\documentclass[aps,prB,reprint,superscriptaddress,onecolumn]{revtex4-1}
\usepackage{amsmath,amssymb}
\usepackage{graphicx}
\usepackage{color}
\draft

\begin{document}
\makeatletter 
\makeatother

\title{Advances toward high-accuracy gigahertz operation of tunable-barrier single-hole pumps in silicon}
\author{Gento Yamahata}
\email{E-mail: gento.yamahata@ntt.com}
\affiliation{NTT Basic Research Laboratories, NTT Corporation, 3-1 Morinosato Wakamiya, Atsugi, Kanagawa 243-0198, Japan}

\author{Akira Fujiwara}
\affiliation{NTT Basic Research Laboratories, NTT Corporation, 3-1 Morinosato Wakamiya, Atsugi, Kanagawa 243-0198, Japan}
\begin{abstract}
Precise and reproducible current generation is key to realize quantum current standards in metrology. A promising candidate is a tunable-barrier single-charge pump, which can accurately transfer single charges one by one with an error rate of less than ppm level. Although several high-accuracy measurements have revealed such a high performance of the pumps, it is necessary to further pursue the possibility of high-precision operation toward reproducible generation of the pumping current in many devices. Here, we investigate in detail a silicon single-hole pumps, which are potentially expected to have a superior performance to single-electron pumps because of a heavy effective mass of holes. Temperature dependence measurements of current generated by the single-hole pump revealed a high energy selectivity of the tunnel barrier, which is a critical parameter to achieve high-accuracy operation. In addition, we applied the dynamic gate compensation technique to the single-hole pump and confirm the further performance improvement. Furthermore, we demonstrate gigahertz operation of a single-hole pump with an estimated lower bound of an error rate of around 0.01 ppm. These results imply a superior capability of single-hole pumps in silicon toward high-accuracy, high-speed, and stable single-charge pumping appropriate for not only metrological applications but also quantum device applications.  
\end{abstract}

\maketitle
\section{Introduction}
Single-charge pumping precisely controls a flow of single charges, enabling us to obtain a clean electric current. Among many potential applications including quantum information processing \cite{SEsouce1,GYnnano, Jcol, Ncol}, single-photon sources \cite{SAWsp}, and quantum sensing \cite{Nathan_samp}, a current standard application is a long-standing goal in the field of metrology \cite{pekola-rev}. Toward this goal, high-accuracy pumping with an error rate of at least less than 0.1 ppm in the nanoampere regime is required. Precise evaluation of the accuracy began in earnest using a GaAs tunable-barrier single-electron (SE) pump about a decade ago \cite{gib1}. After that, many results demonstrating high precision of the tunable-barrier SE pumps were reported using GaAs \cite{PTB-ulca1, Kriss-HR, RobPTB, KrissAIST} and Si \cite{NPL-NTT1, Zhao_pump, LHeSi,Si2GHz} SE pumps, reaching an uncertainty of about 0.2 ppm. In terms of the current level, a nanoampere current was reported using trap-mediated pumping in Si \cite{trap1, trap2} with increasing operating frequency at around 7 GHz. Parallelization of SE pumps is another important pathway to increase the current \cite{PekolaPara, PTBpara, BaePara}. Note that there exist other approaches to obtain a current level in the nanoampere to milliampere regimes \cite{LNECCC,PhaseSlipC,QAHcurrent}. In addition to the accuracy and current level, universality and reproducibility of devices are also important points. Universality of the pump was checked by investigating above many high-accuracy measurements \cite{UnivRev}. In terms of device reproducibility, our Si SE pumps so far show variability of the pumping accuracy \cite{Fcpem22}. The reason of the spread is still under investigation but at least it is necessary to pursue capability of high-accuracy operation for the device to have reproducibility in terms of sub-ppm operation.

In general, the tunable-barrier SE pump in the tunneling regime at a low temperature can be accurately operated with a large charging energy of the quantum dot (QD) in the SE pump and a large energy selectivity of the entrance tunnel barrier characterized by an effective temperature $T_{0}$ \cite{tunable-barrier1, myPRB1}. The former can be achieved by making a small QD and actually our Si devices have typically a large charging energy of about 10 - 20 meV \cite{myapl1, GYnnano,GYcol}. An improvement would be necessary for the energy selectivity of the barrier. This can be achieved by making the barrier with a gentle curvature, which is related to a design of a gate electrode \cite{Fcpem22}. Another way to enhance the energy selectivity is to use a charge carrier with a heavy effective mass. Since a hole is typically heavier than an electron in the case of Si \cite{Sze2}, a single-hole (SH) pump is an attractive candidate for high-accuracy operation. We have previously reported gigahertz operation of SH pumping \cite{myaplhole} but it was high-temperature thermal-hopping regime and the accuracy was on the order of $10^{-3}$ to $10^{-4}$. In addition, a Ge-based SH pump has recently been reported but the operating frequency was limited to 100 MHz and there was no discussion of the accuracy due to charge fluctuations \cite{RosHole}. Therefore, further investigation about SH pumps is highly demanded.

Here, we investigate SH pumps in Si in the tunneling regime and show that the energy selectivity is much better than that in our previous investigation of an SE pump in Si \cite{Nathan_cur}. In addition, we applied gate compensation technique, which has been recently performed using GaAs pumps with two-gate high-frequency operation \cite{FrankComp}, and achieved enhancement of an estimated lower bound of a pumping error rate ($\epsilon _{\mathrm{P}}$) from ppm to ppb levels. Furthermore, we demonstrate 2-GHz SH pumping with $\epsilon _{\mathrm{P}}$ in a deep sub-ppm level.

\section{Device and measurement scheme}
\begin{figure}
\begin{center}
\includegraphics[pagebox=artbox]{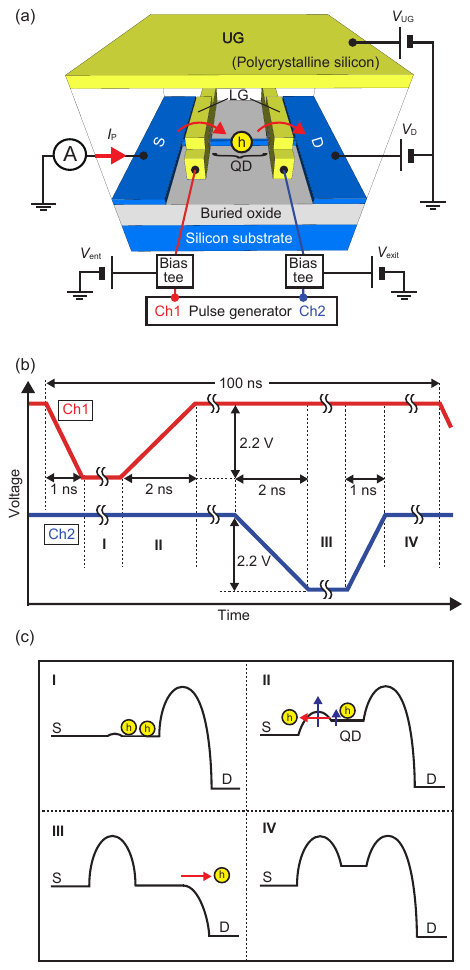}%width=230pt 
\end{center}
\caption{(a) Schematic illustration of the device structure with electrical connections. An yellow circle indicates a single hole. DC voltages ($V_{\mathrm{UG}}$, $V_{\mathrm{D}}$, $V_{\mathrm{ent}}$, $V_{\mathrm{exit}}$) were applied using DC voltage sources (Yokogawa GS200). $I_{\mathrm{P}}$ was converted to voltage using a programmable current amplifier (NF CA5351) and the voltage was measured by a digital multimeter (Keysight 3458A). The pulse generator is Keysight 81160A. (b) One period of voltage pulses applied from the pulse generator. Frequency is 10 MHz. (c) Schematic hole potential diagrams corresponding to regions I - IV in (b). For clarity, the positive and negative directions are reversed from the normal band diagram. At region II, blue arrows indicate the rises in the entrance barrier and QD. At regions II and III, red arrows indicate tunneling of an SH.}
\label{f1}
 \end{figure}

We fabricated a double-layer gate structure on a Si nanowire with a p-type source and drain [Fig. \ref{f1}(a)]. The Si nanowire was patterned using electron beam lithography with dry etching, followed by thermal oxidation to form a gate insulator. The double-layer gate is made of an n-type polycrystalline Si grown by chemical vapor deposition. The upper and lower gates were patterned using electron beam lithograph and optical lithography, respectively, with dry etching. Interlayer oxide between the lower and upper gates was grown by thermal oxidation after the lower-gate formation. The source and drain were doped by boron atoms with ion implantation. The upper gate was used as a mask of the implantation. Aluminum ohmic contacts were finally formed by vacuum deposition. We measured two devices (devices A, B) for this study. The device sizes are summarized in Tab.~\ref{tab1}. We show results of device A in Secs.~III and IV and of device B in Sec.~V.

We set device A in a dilution refrigerator with a base temperature of 50 mK. On the other hand, devices B was measured using a probe station with liquid helium at a stage temperature of around 4 K. For measurements of device A, we connected channels 1 and 2 in a pulse generator to the two lower gates with DC bias $V_{\mathrm{ent}}$ and $V_{\mathrm{exit}}$, respectively. DC voltages $V_{\mathrm{UG}}$ and $V_{\mathrm{D}}$ were also applied to the upper gate and drain, respectively. The DC current $I_{\mathrm{P}}$ through the Si nanowire was measured at the source. For measurements of devices B, we used sinusoidal-signal generator instead of the channel 1 of the pulse generator and the channel 2 of the pulse generator was disconnected. Other connections were the same as those for device A.

Figure \ref{f1}(b) schematically illustrates the outputs of the pulse generator. The frequency $f$ of the voltage pulses is 10 MHz. Note that the rise time of 2 ns in the channel 1 is roughly equivalent to 250-MHz operation with an application of a sinusoidal signal. For device A, we used turnstile operation to transfer SHs \cite{fturn1}, in which sufficiently high drain bias was applied to have unidirectional transfer, because it can be easily extended to the gate compensation technique discussed in Sec.~IV. The hole potential diagrams corresponding to regions I, II, III, and IV in Fig.~\ref{f1}(b) are shown in Fig.~\ref{f1}(c). In region I, some holes are loaded in the region between the two lower gates. In region II, the entrance barrier is raised. A QD is formed between the entrance and exit barriers. During this process, the QD potential is also raised because of the capacitive coupling between the entrance lower gate and QD. When the QD electrochemical potential exceeds the Fermi level of the source, some holes escape back to the source and finally an SH can be captured by the QD because the escape rate to the source eventually becomes negligibly slow. We refer to this process as a dynamic capture process and it was theoretically formulated using the decay cascade model \cite{kaestner1}. After the rise of the entrance barrier, the exit barrier is lowered and it is the lowest in region III. During this process, the captured SH is ejected to the drain. Then, the exit barrier is raised but no additional SH is captured because of the large drain bias. When the number of the transferred SH is $n$, the output current becomes $nef$, where $e$ is the elementally charge. For device B, the pulse voltage of the channel 1 is replaced to a sinusoidal signal. Without lowering exit barrier, a captured SH can be ejected to the drain because the QD potential becomes sufficiently high \cite{frat1}. The purpose of measurements of the device B is to check capability of high-frequency operation shown in Sec.~V.

\section{High energy selectivity of tunnel barrier: device A}
\begin{figure}
\begin{center}
\includegraphics[pagebox=artbox]{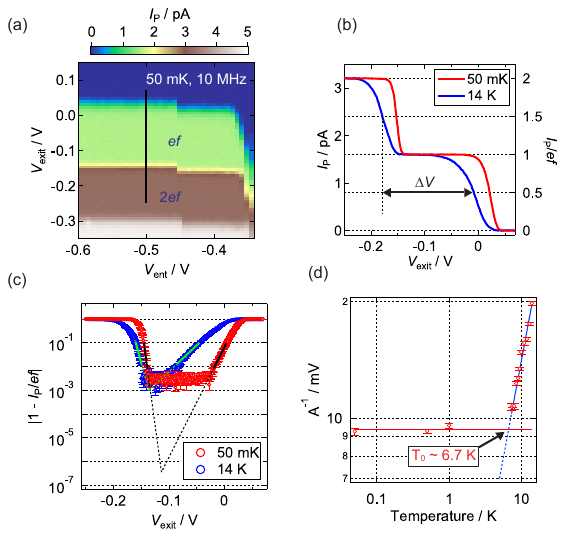}%width=230pt 
\end{center}
\caption{(a) $I_{\mathrm{P}}$ as a function of $V_{\mathrm{ent}}$ and $V_{\mathrm{exit}}$ at 50 mK, where $V_{\mathrm{UG}}=-2.5$ V and $V_{\mathrm{D}}=-0.5$ V. Delay between the channels 1 and 2 is 50 ns. (b), (c) $I_{\mathrm{P}}$ and $|1-I_{\mathrm{P}}/ef|$ as a function of $V_{\mathrm{exit}}$ at 50 mK and 14 K, where $V_{\mathrm{UG}}=-2.5$ V, $V_{\mathrm{D}}=-0.5$ V, and $V_{\mathrm{ent}}=-0.5$ V. Error bars in (c) are standard deviations of the measurement results. (d) Inverse of the slope extracted from exponential fits to the right side of $|1-I_{\mathrm{P}}/ef|$ as a function of temperature. A red horizontal line is an average of three points from the left. A blue tilted line is a linear fit to the data and a dashed blue line is an extension of it. An intersection of them is defined as $T_{0}$.}
\label{f2}
 \end{figure}

Figure~\ref{f2}(a) shows $I_{\mathrm{P}}$ as a function of $V_{\mathrm{ent}}$ and $V_{\mathrm{exit}}$ for device A at 50 mK. $I_{\mathrm{P}}$ along a black line in Fig.~\ref{f2}(a) is plotted as a red line in Fig.~\ref{f2}(b), where the current increase is determined by the dynamic capture process because of no $V_{\mathrm{ent}}$ dependence in Fig.~\ref{f2}(a). To evaluate the accuracy of the $ef$ plateau, we plot $|1-I_{\mathrm{P}}/ef|$ [red circles in Fig.~\ref{f2}(c)], which indicates the relative deviation of the current plateau from an expected value of $ef$. Because of limitation of the measurement system, we measured the current with an uncertainty of the order of $10^{-3}$. From this kind of data, it is valuable to extract $\epsilon _{\mathrm{P}}$ by extending exponential lines of the relative deviation \cite{LHeSi, GYgval}. Although this method does not reveals a true accuracy of the pump and high-accuracy measurements are definitely necessary, it would be useful information because we can conveniently select a pump with a high possibility of high-accuracy operation and so far this extension looks reasonable on the sub-ppm level \cite{Si2GHz}. A black line on the right in Fig.~\ref{f2}(c) is a fit to the data with a formula of $\mathrm{exp}\{A (V_{\mathrm{exit}}-V_{0})\}$ [see Eq.~(\ref{tay})], where $A$ is the slope. A black line on the left in Fig.~\ref{f2}(c) is a phenomenological fit by a similar exponential function. We use $A$ extracted from the former fit in the following analysis. The intersection of two black dashed lines extending the two exponential fits indicates $\epsilon _{\mathrm{P}}$, which is about 0.69 ppm (twice the intersection value) in this specific device.

To more quantitatively understand $\epsilon _{\mathrm{P}}$, it is helpful to investigate a figure of merit parameter $\delta = (1+1/g)E_{\mathrm{C}}/(kT_{0})$ determining the pumping accuracy at a low temperature \cite{tunable-barrier1,GYgval}, where $E_{\mathrm{C}}$ is the charging energy of the QD, $k$ is the Boltzmann constant, and $g$ is a capacitive coupling parameter [see Eq.~(\ref{gdef})]. Note that a typical $g$ value of our devices is on the order of 10 \cite{Fcpem22}. To extract $T_{0}$, we investigated temperature dependence of the inverse of the slope ($A^{-1}$) because it is proportional to $T_{0}$ [see Eq.~(\ref{Dec1})] at a low temperature but at a high temperature it is proportional to $T$. Therefore, the temperature at which $A^{-1}$ begins to saturate corresponds to $T_{0}$. A blue line in Fig.~\ref{f2}(b) and blue circles in Fig.~\ref{f2}(c) are $I_{\mathrm{P}}$ and $|1-I_{\mathrm{P}}/ef|$ at 14 K, respectively. Similar to the 50-mK data, we fit the both side of $|1-I_{\mathrm{P}}/ef|$ and extract $A^{-1}$ at the right side. Note that $\epsilon _{\mathrm{P}}$ at 14 K is $1.9\times 10^{3}$ ppm. Figure~\ref{f2}(d) shows $A^{-1}$ as a function of temperature. As expected, $A^{-1}$ is constant in the low temperature regime but at around 10 K it linearly increases. A blue line in Fig.~\ref{f2}(d) is a linear fit to $A^{-1}$. A red horizontal line is a mean value of the low temperature data at 1, 0.5, and 0.05 K. We estimate $T_{0}$ from the intersection point between the extension of the blue line (blue dashed line) and the red line, resulting in $T_{0}\sim 6.7$ K. We additionally evaluate $E_{\mathrm{C}}$ from the 14-K data shown in Fig.~\ref{f2}(b). From the value of $A$ and the spacing of plateau $\Delta V$ of about 170 mV, we extract $(1+1/g)E_{\mathrm{C}} = kTA\Delta V \sim 10$ meV. This value is similar to that of the previous SH pump ($E_{\mathrm{C}}\sim 14$ meV) \cite{myaplhole}. From the extracted values, we obtain $\delta$ of about $17$ at a low temperature.

In the previous estimation of our Si SE pump \cite{Nathan_cur}, $(1+1/g)E_{\mathrm{C}}\sim 17.8$ meV and $T_{0} = 17$ K. Although charging energy was larger in this case, the large $T_{0}$ leads to $\delta \sim 12$. This indicates that the low $T_{0}$ in the SH pump leads to the good characteristics. Theoretically, it is likely to be attributed to the difference of the effective mass $m$ between electron ($m_{\mathrm{e}}$: typically, $0.19m_{0}$) and hole ($m_{\mathrm{h}}$: typically, $0.49m_{0}$), where $m_{0}$ is the bare electron mass \cite{Sze2,myaplhole}. Assuming a parabolic potential barrier characterized by $-m\omega ^{2}x^{2}/2$, barrier curvature $S_{0}$ at $x=0$ is $m\omega ^{2}$. Therefore, $\omega = \sqrt{S_{0}/m}$. In this case, $T_{0} = \hbar \omega/2 \pi k \propto 1/\sqrt{m}$ \cite{myPRB1}. Therefore, $T_{0}$ of a hole could be $\sqrt{m_{\mathrm{e}}/m_{\mathrm{h}}}\sim 0.6$ times larger than that of an electron. When $E_{\mathrm{C}}$ is independent of the effective mass (i.e., a self capacitance dominates $E_{\mathrm{C}}$ \cite{myapl1}), this directly leads to the enhancement of $\delta$ of about 1.6. This enhancement corresponds to more than two order magnitude enhancement of $\epsilon _{\mathrm{P}}$. 

%Since the potential shape of the SH pump is not same as the previous SE pump, the above result is not direct comparison but the low value of $T_{0}$ of the SH pump is marked characteristics to show the advantage of the SH pumping.

\section{Dynamic gate compensation: device A}
$\epsilon _{\mathrm{P}}\sim 0.69$ ppm of device A is a relatively good value but is not sufficient for the metrological applications. Here, we use dynamic gate compensation technique proposed in Ref.~\cite{FrankComp} to the Si SH pump to show the universality of the technique and potential of our device. The concept of the compensation is to suppress the change in the potential of the QD in the dynamic capture process by additionally changing the signal applied to the exit gate. In our case, we changed the delay time of the voltage pulses in the channel 2 to match the voltage rise in the channel 1 with the voltage fall with a time duration of $\tau _{\mathrm{c}}$ in the channel 2 [Fig.~\ref{f3}(a)]. At region II, the QD potential is additionally lowered by the signal applied to the exit gate [Fig.~\ref{f3}(b)]. This leads to the change in the dynamic capture probability.

Following the method described in Ref.~\cite{GYgval}, we include the compensation effect to the dynamic capture probability as 
\begin{equation}
 P_{1}=
 \mathrm{exp}
 \left [
 -\mathrm{exp}
 \left \{
-\frac{
\alpha_{\mathrm{C}}(r)
\left (
rV_{\mathrm{ent}}
+V_{\mathrm{exit}}
\right )
+(1+1/g_{\mathrm{c}})E_{\mathrm{C}}
}
{kT_{0}}
+\mathrm{const.}
\right \}
\right ],
 \label{Dec1m}
 \end{equation}
where $r$ is the ratio of the slopes of the high-frequency signals applied to the entrance and exit gates (see Appendix A), $\alpha _{\mathrm{C}}(r)$ is defined as Eq.~(\ref{alpfit}), and $g_{\mathrm{c}}$ is the effective capacitive coupling parameter with the compensation effect given as Eq.~(\ref{gcdef}). In addition, a figure of merit parameter $\delta_{\mathrm{c}}$ determining the accuracy of the dynamic capture can be written as (see Appendix A),
\begin{equation}
\delta _{\mathrm{c}} = \left (
1 + \frac{1}{g_{\mathrm{c}}}
\right )
\frac{E_{\mathrm{C}}}
{kT_{0}}.
\end{equation}
Since $g_{\mathrm{c}}$ decreases with increasing $r$ [Fig.~\ref{f3}(c)], $\delta _{\mathrm{c}}$ (i.e., $\epsilon _{\mathrm{P}}$) increases (decreases) with increasing $r$ [Fig.~\ref{f3}(d)]. Note that these discussions are identical to what were discussed in Ref.~\cite{FrankComp} but our purpose here is to clearly show a relation between the compensation technique to the discussion of the effect of $g$ describing in Ref.~\cite{GYgval}, which would helps understanding of the mechanism of the compensation effect.
%%%%

We measured $I_{\mathrm{P}}$ at $\tau_{\mathrm{c}}=2$ ns as a function of $V_{\mathrm{ent}}$ and $V_{\mathrm{exit}}$ [Fig.~\ref{f4}(a)]. Since the slope of the line with the same current level dominated by the dynamic capture process determines only by $r$ [see Eq.~(\ref{Dec1m})], we perform a linear fit to the line [red line in Fig.~\ref{f4}(a)], resulting in $r$ of 0.62. The current plateau along a black line in Fig.~\ref{f4}(a) is plotted in Fig.~\ref{f4}(b). Figure \ref{f4}(c) is $|1-I_{\mathrm{P}}/ef|$ for this case. Similar to Fig.~\ref{f2}(c), we extracted the slope of the current and plot the inverse of the slope as a function of $r$ extracted from pumping current maps [Fig.~\ref{f4}(d)]. Note that $r$ was changed by changing $\tau _{\mathrm{c}}$. A blue line in Fig.~\ref{f4}(d) is a fit to $A^{-1}$ using Eq.~(\ref{alpfit}), yielding $\alpha _{\mathrm{B}}=0.16$ eV/V and $\alpha _{\mathrm{QDc}}^{g}=0.079$ eV/V, where $\alpha _{\mathrm{QDc}}^{g}=(1+1/g)\alpha _{\mathrm{QDc}}$, and $\alpha _{\mathrm{B}}$ and $\alpha _{\mathrm{QDc}}$ are the voltage-to-energy conversion factors between the entrance gate and entrance barrier and between the exit gate and QD, respectively. This good fit indicates effectiveness of the gate compensation technique. Note that slight fluctuation of the value of the slope could be related to potential fluctuation in the QD. We also extracted $\epsilon _{\mathrm{P}}$ by extending the exponential fits in Fig.~\ref{f4}(c). With increasing $r$, $\epsilon _{\mathrm{P}}$ decreases from ppm to ppb levels [inset in Fig.~\ref{f4}(d)], which roughly corresponds to the simplified expectation shown in Fig.~\ref{f3}(d) [the extracted conversion factors are used in Figs.~\ref{f3}(c) and (d)]. This accuracy level is sufficient for the metrological application.

\section{Gigahertz operation: device B}
\begin{figure}
\begin{center}
\includegraphics[pagebox=artbox]{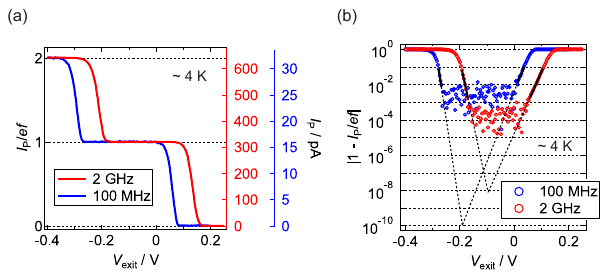}%width=230pt 
\end{center}
\caption{(a) $I_{\mathrm{P}}$ as a function of $V_{\mathrm{exit}}$ at around 4 K for device B. In this measurements, high-frequency sinusoidal signal was applied to the entrance gate by the signal generator (Keysight E8257D) with a power of $P$. In 100-MHz data, $V_{\mathrm{UG}}=-2.25$ V, $V_{\mathrm{D}}=0$ V, $V_{\mathrm{ent}}=0.2$ V, and $P=7$ dBm. In 2-GHz data, $V_{\mathrm{UG}}=-2.25$ V, $V_{\mathrm{D}}=0$ V, and  $V_{\mathrm{ent}}=0.25$ V, and $P=9$ dBm. (b) $|1-I_{\mathrm{P}}/ef|$ calculated from data in (a). Black lines are exponential fits.}
\label{f5}
 \end{figure}

In another device, we measured frequency dependence of the current plateau of the SH pumping. Figure \ref{f5}(a) shows $I_{\mathrm{P}}$ as a function of $V_{\mathrm{exit}}$ for device B at 100 MHz and 2 GHz at around 4 K. Corresponding $|1-I_{\mathrm{P}}/ef|$ is plotted in Fig.~\ref{f5}(b). Similar to the above analysis, we perform an exponential fit to the data and extract $\epsilon _{\mathrm{P}}$. In this specific device, the error rate is on the order of 0.01 ppm for the 2-GHz operation. This also indicates potential of the SH pumping. We also observe the increase in $\epsilon _{\mathrm{P}}$ with increasing frequency. This is a general trend of single-charge pumps \cite{gib1, Seo1, myaplhole, NPL-NTT1} possibly due to nonadiabatic excitation of a hole confined in the QD suring the dynamic QD movement \cite{kataoka1} and it should be further investigated in future works.

\section{Conclusions}
We have observed a high energy selectivity of the entrance barrier (low $T_{0}$) of a Si SH pump. This superior property is important for the high-accuracy operation with high yield. By using the gate compensation technique, we confirmed that the pump performance was further improved to the level suitable for the metrological application. Furthermore, we have demonstrated 2-GHz SH pumping in another SH pump with an estimated lower bound of error rate of about 0.01 ppm. These results would indicate high potential of SH pumps. In future work, it is necessary to check variability of $T_{0}$ among many devices with an identical design. In addition, pumping characteristics of an SH and SE should be compared with the same QD by making N-type and P-type ohmic contacts simultaneously \cite{nob1}.

\begin{acknowledgements}
We thank T. Shimizu, F. Hohls, and V. Kashcheyevs for fruitful discussions. This work was partly supported by JSPS KAKENHI Grant Number JP18H05258.
\end{acknowledgements}

\section*{Data availability}
The data that support the findings of the study are available from the corresponding author upon reasonable request.

\appendix \section{Formulation of dynamic gate compensation}
\begin{figure}
\begin{center}
\includegraphics[pagebox=artbox]{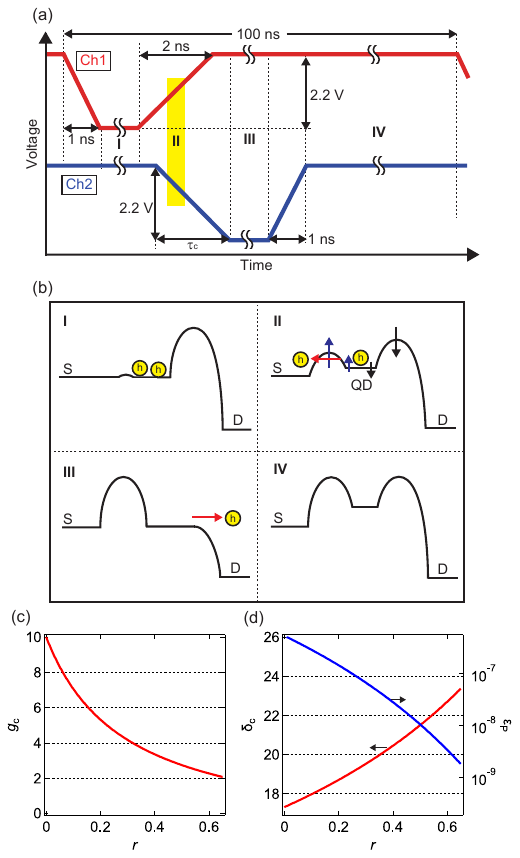}%width=230pt 
\end{center}
\caption{(a) One period of voltage pulses applied from the pulse generator, where there is an overlap between the voltage rise of the channel 1 and the voltage fall of the channel 2. (b) Schematic hole potential diagrams corresponding to region I - IV in (a). For clarity, the positive and negative directions are reversed from the normal band diagram. At region II, blue (black) arrows indicate the rises (falls, i.e., the compensation effect) in the entrance (exit) barrier and QD. At regions II and III, red arrows indicate tunneling of an SH. (c) Calculated $g_{\mathrm{c}}$ as a function of $r$. Here, we use $\alpha _{\mathrm{B}}=0.16$ eV/V and $\alpha_{\mathrm{QD}}^{g}=0.079$ eV/V extracted from the fit in Fig.~\ref{f4}(d). We assume that $g=10$. (d) Calculated $\delta _{\mathrm{c}}$ using $g_{\mathrm{c}}$ in (c) (left) and $\epsilon _{\mathrm{P}}=6.2\mathrm{exp}(-0.94\delta_{\mathrm{c}})$ \cite{fbook} (right) as a function of $r$, where $E_{\mathrm{C}}=10$ meV and $T_{0}=6.7$ K.}
\label{f3}
 \end{figure}

\begin{figure}
\begin{center}
\includegraphics[pagebox=artbox]{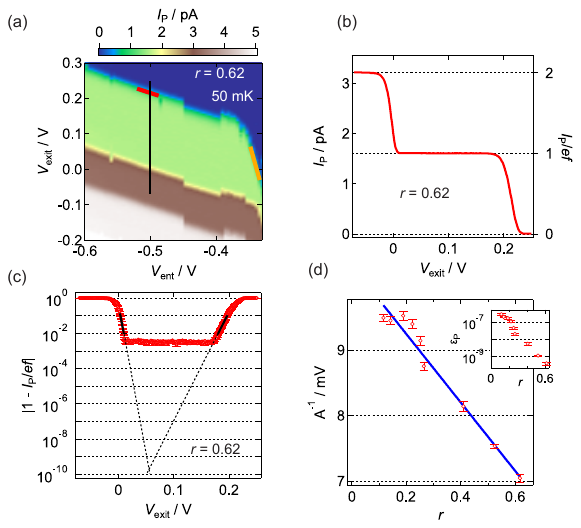}%width=230pt 
\end{center}
\caption{(a) $I_{\mathrm{P}}$ as a function of $V_{\mathrm{ent}}$ and $V_{\mathrm{exit}}$ at 50 mK, where $V_{\mathrm{UG}}=-2.5$ V, $V_{\mathrm{D}}=-0.5$ V, and $\tau _{\mathrm{c}}=2$ ns. Delay between the channels 1 and 2 is 25.5 ns. (b), (c) $I_{\mathrm{P}}$ and $|1-I_{\mathrm{P}}/ef|$ as a function of $V_{\mathrm{exit}}$ at 50 mK, where $V_{\mathrm{UG}}=-2.5$ V, $V_{\mathrm{D}}=-0.5$ V, $V_{\mathrm{ent}}=-0.5$ V, and $\tau _{\mathrm{c}}=2$ ns. Delay between the channels 1 and 2 is 25.5 ns. Error bars in (c) are standard deviations of the measurement results. (d) Inverse of the slope extracted from exponential fits to the right side of $|1-I_{\mathrm{P}}/ef|$ as a function of $r$. A blue line is a fit to the data using Eq.~(\ref{alpfit}). Inset: the estimated lower bound of the error rate extracted from the intersection of the exponential fits as in (c) as a function of $r$.}
\label{f4}
 \end{figure}

We formulate the gate-compensation effect for an SH pumping [region II in Fig.~\ref{f3}(a)] using a capacitive coupling parameter $g$ and its extended version $g_{\mathrm{c}}$ following the method described in Ref. \cite{GYgval}. For simplicity, we use the constant interaction model \cite{taruqd1, dqd1} with a single level. In this case, the potential energy of the QD with one electron and the height of the entrance barrier can be calculated as 
\begin{align}
E_{\mathrm{QD}}(t)&=
\alpha_{\mathrm{QD}}
\left (
V_{\mathrm{ent}}+\gamma t
\right )
+\alpha_{\mathrm{QDc}}
\left (
V_{\mathrm{exit}}-r\gamma t
\right )
-E_{\mathrm{C}},\\
%E_{\mathrm{D},2}(t)&=
%-\alpha_{\mathrm{D}}
%\left (
%V_{\mathrm{ent}}-t/\tau _{\mathrm{p}}
%\right )
%-\alpha_{\mathrm{CD}}
%\left (
%V_{\mathrm{exit}}+t/\tau _{\mathrm{c}}
%\right ),\\
E_{\mathrm{B}}(t)&=
\alpha_{\mathrm{B}}
\left (
V_{\mathrm{ent}}+\gamma t
\right )
+\alpha_{\mathrm{Bc}}
\left (
V_{\mathrm{exit}}-r\gamma t
\right ),
\end{align}
respectively, where we set $E_{\mathrm{QD}}(0)=-E_{\mathrm{C}}$ and $E_{\mathrm{B}}(0)=0$ when $V_{\mathrm{ent}}=V_{\mathrm{exit}}=0$, $E_{\mathrm{C}}$ is the charging energy of the QD, $\alpha_{\mathrm{QD(c)}}$ and $\alpha_{\mathrm{B(c)}}$ are voltage-to-energy conversion factors between the entrance (exit) gate and QD and between the entrance (exit) gate and entrance barrier, respectively, we assume a linear ramp of the voltage pulses with a coefficient of $\gamma$, and $r$ is the ratio of the slopes of the voltage pulses at the entrance gate to that of the exit gates. Note that the signs preceding the voltage-to-energy conversion factors are positive because our focus is an SH pumping. The parameter $g$ ($g_{\mathrm{c}}$) is defined as a ratio between the QD potential change and the change in the barrier height with respect to the QD potential during the rise of the entrance barrier without (with) the compensation, given as
\begin{align}
g &\equiv \frac{\alpha _{\mathrm{QD}}}{\alpha _{\mathrm{B}}-\alpha _{\mathrm{QD}}},\label{gdef} \\
g_{\mathrm{c}} &\equiv \frac{\alpha _{\mathrm{QD}}-r\alpha _{\mathrm{QDc}}}{(\alpha _{\mathrm{B}}-\alpha _{\mathrm{QD}})-r(\alpha _{\mathrm{Bc}}-\alpha _{\mathrm{QDc}})}.
\label{gcdef}
\end{align}
Then, the tunnel rate $\Gamma _{1}(t)$ of a hole between the QD and source is calculated as
\begin{align}
\Gamma _{1}(t) &= \Gamma _{0}
\mathrm{exp}
\left \{
-\frac{E_{\mathrm{B}}(t)-E_{\mathrm{QD}}(t)}{kT_{0}}
\right \} \label{gam1}
\\
&= \Gamma_{v}(V_{\mathrm{ent}}, V_{\mathrm{exit}}, E_{\mathrm{C}})\mathrm{exp}(-\Gamma_{\mathrm{inc}}t),
%\\
%\Gamma _{2}(t) &= 2\Gamma _{0}
%\mathrm{exp}
%\left \{
%-\frac{E_{\mathrm{B}}(t)-E_{\mathrm{D},2}(t)}{kT_{0}}
%\right \}.
%\label{gam2}
\end{align}
where
\begin{align}
\Gamma_{v}(V_{\mathrm{ent}}, V_{\mathrm{exit}}, E_{\mathrm{C}}) &= \Gamma _{0}
\mathrm{exp}
\left \{
-\frac{(\alpha_{\mathrm{B}}-\alpha_{\mathrm{QD}})V_{\mathrm{ent}}+(\alpha_{\mathrm{Bc}}-\alpha_{\mathrm{QDc}})V_{\mathrm{exit}}+E_{\mathrm{C}})}{kT_{0}}
\right \},
\\
\Gamma _{\mathrm{inc}}&= \frac{\gamma (\alpha_{\mathrm{QD}}-r\alpha_{\mathrm{QDc}})}{kT_{0}g_{\mathrm{c}}},
%\\
%\Gamma _{2}(t) &= 2\Gamma _{0}
%\mathrm{exp}
%\left \{
%-\frac{E_{\mathrm{B}}(t)-E_{\mathrm{D},2}(t)}{kT_{0}}
%\right \}.
%\label{gam2}
\end{align}
and $\Gamma_{0}$ is a constant. Note that $T_{0}$ is replaced with $T$ in the high-temperature thermal-hopping regime. A time $t_{1}^{\mathrm{c}}$ when the detailed balance breaks down is defined as $\int_{t_{1}^{\mathrm{c}}}^{\infty}\Gamma_{1}(t)dt=1$, leading to $t_{1}^{\mathrm{c}}=\mathrm{ln}\{ \Gamma_{v}(V_{\mathrm{ent}}, V_{\mathrm{exit}}, E_{\mathrm{C}})/\Gamma _{\mathrm{inc}} \}/\Gamma _{\mathrm{inc}}$. Then, we obtain
\begin{equation}
E_{\mathrm{QD}}^{\mathrm{c}}\equiv E_{\mathrm{QD}}(t_{1}^{\mathrm{c}})=
\left \{
(1+g_{\mathrm{c}})\alpha_{\mathrm{QDc}}
-g_{\mathrm{c}}
\alpha_{\mathrm{Bc}}
\right \}
\left (
rV_{\mathrm{ent}}
+V_{\mathrm{exit}}
\right )
-(1+g_{\mathrm{c}})E_{\mathrm{c}} + \mathrm{const.}
\end{equation}
This equation is equivalent to Eq.~(14) of Ref.~\cite{GYgval} when $r=0$ (no compensation). The SH dynamics is governed by the master equation [Eqs.~(4)-(6) in Ref.~\cite{GYgval}] and the steady state is expressed as [see Eqs.~(8) and (13) in Ref.~\cite{GYgval}]
\begin{align}
P_{1}&=\int_{0}^{\infty}\mathrm{exp}
\left [
-\mathrm{exp}
\left \{
-\Gamma _{\mathrm{inc}}(t-t_{1}^{\mathrm{c}})
\right \}
\right ]
\left [
-\frac{df\{ E_{\mathrm{QD}}(t) \}}{dt}
\right ]
dt\\
&=\int _{E_{0}}^{\infty}\mathrm{exp}
 \left \{
 -\mathrm{exp}
 \left (
 -\frac{E_{\mathrm{QD}}-E_{\mathrm{QD}}^{\mathrm{c}}}
 {g_{\mathrm{c}}kT_{0}}
 \right )
 \right \}
 \left \{
 -\frac{df\left ( E_{\mathrm{QD}}\right )}{dE_{\mathrm{QD}}}
 \right \}
 dE_{\mathrm{QD}},
\label{inteq}
\end{align}
where $f(x)=\{ 1+\mathrm{exp}(x/kT) \}^{-1}$ is the Fermi function, we use the relation of $\gamma (\alpha _{\mathrm{QD}}-r\alpha _{\mathrm{QDc}})(t-t_{1}^{\mathrm{c}})=E_{\mathrm{QD}}(t)-E_{\mathrm{QD}}(t_{1}^{\mathrm{c}})$, and $E_{\mathrm{0}}=E_{\mathrm{QD}}(0)$. From this equation, the probability distribution for the decay cascade model can be calculated as [see Eq.~(16) in Ref.~\cite{GYgval}]
\begin{align}
 P_{1}&=
  \mathrm{exp}
 \left \{
 -\mathrm{exp}
 \left (
 \frac{E_{\mathrm{QD}}^{\mathrm{c}}}
 {g_{\mathrm{c}}kT_{0}}
 \right )
 \right \}\\
 &=\mathrm{exp}
 \left [
 -\mathrm{exp}
 \left \{
\frac{
\alpha_{\mathrm{C}}(r)
\left (
rV_{\mathrm{ent}}
+V_{\mathrm{exit}}
\right )
-(1+1/g_{\mathrm{c}})E_{\mathrm{C}}
}
{kT_{0}}
+\mathrm{const.}
\right \}
\right ],
 \label{Dec1}
 \end{align}
where
\begin{align}
\alpha_{\mathrm{C}}(r)&=
\left (1+\frac{1}{g_{\mathrm{c}}}\right )
\alpha_{\mathrm{QDc}}-\alpha_{\mathrm{Bc}}\\
&=
\left \{
\left (1+\frac{1}{g}\right )
\alpha_{\mathrm{QDc}}
-\alpha_{\mathrm{Bc}}
\right \}
\left  (
1-
\frac
{\alpha_{\mathrm{QDc}}}
{\alpha_{\mathrm{QD}}}
r
\right )^{-1}
\label{alpC}
\\
&=\left (
\alpha_{\mathrm{QDc}}^{g}
-\alpha_{\mathrm{Bc}}
\right )
\left  (
1-
\frac
{\alpha_{\mathrm{QDc}}^{g}}
{\alpha_{\mathrm{B}}}
r
\right )^{-1},
\label{alpfit}
\end{align}
and $\alpha_{\mathrm{QDc}}^{g}=(1+1/g)\alpha_{\mathrm{QDc}}$. Equation~(\ref{alpC}) is identical to geometrically derived Eq.~(4) in Ref.~\cite{FrankComp}. For the fit to $A^{-1}$ in Fig.~\ref{f4}(d), we used Eq.~(\ref{alpfit}). For estimation of $\alpha _{\mathrm{Bc}}$, we used the slope $\Delta$ of the loading process [region I in Fig.~\ref{f3}(a)] indicated by an orange line in Fig.~\ref{f4}(a). Since this loading process is dominated by the entrance barrier height, $\alpha _{\mathrm{Bc}}=\alpha _{\mathrm{B}}/|\Delta|$ \cite{GYnnano}. The orange line in Fig.~\ref{f4}(a) is a linear fit to a contour line with a current level of 1 pA, yielding $\Delta = -6.8$. For the evaluation of $|1-I_{\mathrm{P}}/ef|$, we calculate $1-P_{1}$ as
\begin{align}
1-P_{1}&=1-
\mathrm{exp} \left [
-\mathrm{exp} \left \{
A(V_{\mathrm{exit}}-V_{0})
\right \}
\right ]\\
&\sim \mathrm{exp} \left \{
A(V_{\mathrm{exit}}-V_{0})
\right \},
\label{tay}
\end{align}
where we assume that $\mathrm{exp}\{ A(V_{\mathrm{exit}}-V_{0}) \} \ll 1$. Furthermore, since Eq.~(\ref{Dec1}) is identical to Eq.~(16) in Ref.~\cite{GYgval} when $g_{\mathrm{c}}$ is replaced to $g$, a figure of merit parameter $\delta _{\mathrm{c}}$ when the compensation is applied can be written as [see Eq.~(20) in Ref.~\cite{GYgval}]
\begin{align}
\delta _{\mathrm{c}} &= \left (
1 + \frac{1}{g_{\mathrm{c}}}
\right )
\frac{E_{\mathrm{C}}}
{kT_{0}}.
%&=
%\left (
%1-
%\frac
%{\alpha_{\mathrm{CB}}}
%{\alpha_{\mathrm{B}}}
%r
%\right )
%\left (
%1-
%\frac
%{\alpha_{\mathrm{CD}}^{g}}
%{\alpha_{\mathrm{B}}}
%r
%\right )^{-1}\delta.
\end{align}
%\bibliography{bunken.bib}

\clearpage

\begin{table}%[htb]
 \begin{center}
    \caption{Summary of the device sizes.}
    \begin{tabular}{ccccc} \hline \hline
     Device
     & Wire thickness
      & Wire width
      & Gate space
      & Gate length \\ \hline
       A
      & 15 nm
      & 20 nm
      & 80 nm
      &  40 nm  \\
       B
     & 15 nm
     & 20 nm
     & 80 nm
     & 10 nm \\ \hline \hline
  %   C
  %   & 15 nm
  %    & 30 nm
  %    & 80 nm
  %   & 10 nm\\ \hline \hline
    \end{tabular}
    \label{tab1}
  \end{center}
\end{table}
\clearpage

\end{document}